\documentstyle[12pt,graphicx]{article}
\title{ Relic Gravitational Waves and the Evolution of the
Universe
      }
\author{\small  Wen  Zhao \\
        \small  Astrophysics Center \\
        \small  University of Science and Technology of China \\
        \small  Hefei, Anhui, China }
 \date{}

\begin{document}
\maketitle
\baselineskip=19truept
\def\vek{\vec{k}}

\newcommand{\be}{\begin{equation}}
\newcommand{\ee}{\end{equation}}
\newcommand{\ba}{\begin{eqnarray}}
\newcommand{\ea}{\end{eqnarray}}

\sf
\small

\begin{center}
\Large  Abstract
\end{center}
\begin{quote}
 {\small
 In the inflation models, the relic gravitational waves (RGW)
 generated in the
 inflation stage, and evolved in the Universe until now. In the
 different cosmological evolution models, one can get different gravitational
 waves power spectrum. In this paper, we give a simple formula to
 estimate this spectrum in a general cosmological model. From this
 formula, one can easily find the relation between this power spectrum and
 the cosmological evolution models. The spectrum includes all the
 information about the evolution of the scale factor $a$ from inflation to now.
 So RGW is a more clear fossil, which records the cosmological evolution
 information, and be a useful complementarity for another fossil$-$Cosmic
 Microwave Background Radiation (CMB). }
\end{quote}

PACS numbers:    98.80.-k~  98.80.Es~ 04.30.-w~  04.62.+v


e-mail: wzhao7@mail.ustc.edu.cn

\baselineskip=19truept

\begin{center}
{\em\Large }
\end{center}

In the inflation expansion cosmic models, the stochastic
background of relic gravitational waves can generate from the
quantum fluctuate, which is important in cosmology and has been
extensively studied in the past\cite{muk}. The power spectrum of
it, which is observed today, depends not only on the details of
the early stage of inflationary expansion, but also on the
expansion behavior of the subsequent stages\cite{allen,zhang},
especially, some stages as the pre-big bang stage, the reheating
stage, the cosmological QCD transition, the $e^+e^-$ annihilation
process, the accelerating expansion process, which we did not know
very clearly in physics until now. Their information has also been
recorded in the RGW power spectrum. Most of these stages occurred
before the CMB photon decoupled, which makes it difficult to study
them from CMB. So research on RGW power spectrum is nearly the
only very way to study this unknown process of the Universe. In
this paper, I will study this power spectrum in a general
cosmological model, and find that there is a very simple formula
to estimate it. From this formula. one can easily find the
relation between the spectrum and the evolution of the Universe.
Especially, the spectrum index directly relates to the state
equation of matter which dominated the evolution of the Universe
at some stage. This gives a simple way to estimate the cosmic
evolution process from the RGW power spectrum.

In the spatially flat Robertson-Walker spacetime, the metric which
includes the gravitational waves is\
 \be ds^2=a^2(\tau) [
 d\tau^2-(\delta_{ij}+h_{ij})dx^idx^j ],
 \ee
where $\tau$ is the conformal time, the perturbations of spacetime
$h_{ij}$ is a $3\times 3$ symmetric matrix. The gravitational wave
field is the tensorial portion of $h_{ij}$, which is
transverse-traceless $\partial_i h^{ij}=0$,$\delta^{ij}h_{ij}=0.$
Since  the relic gravitational waves are very weak, $|h_{ij}|  \ll
1$, so one needs just study the linearized field equation:
 \be
 \partial_{\mu}(\sqrt{-g}\partial^{\mu}h_{ij}({\bf{x}} ,\tau))=0 .
 \ee

In quantum theory of gravitational waves, the field $h_{ij}$ is a
field operator, which is written as a sum of the plane wave
Fourier modes
 \be
 h_{ij}({\bf{x}},\tau) = \frac{  \sqrt{16\pi}l_{Pl}   }{(2\pi)^{\frac{3}{2}}} \sum_{\lambda=1}^2
 \int_{-\infty}^{\infty} d^3{\bf k} \epsilon^{(\lambda)}_{ij}({\bf
 k}) \frac{1}{\sqrt{2k}} [ a_k^{(\lambda)}  h^{(\lambda)}_k(\tau)
 e^{i\bf k\cdot x} + a_k^{\dagger(\lambda)} h^{(\lambda)
 *}_k(\tau)e^{-i\bf k\cdot x} ],
 \ee
where $l_{Pl} = \sqrt{G}$ is the Planck's length, the two
polarizations $\epsilon^{(\lambda)}_{ij}(\bf k)$, $\lambda =1, 2$,
are symmetric and transverse-traceless $
k^i\epsilon^{(\lambda)}_{ij}({\bf k})=0$, $ \delta^{ij}
\epsilon^{(\lambda)}_{ij}({\bf k})=0  $, and satisfy the
conditions $  \epsilon ^{(\lambda)ij}({\bf k})
\epsilon^{(\lambda')}_{ij}({\bf k}) =  2\delta_{\lambda\lambda'}
$, and $\epsilon^{(\lambda)}_{ij}({\bf -k}) =
\epsilon^{(\lambda)}_{ij}({\bf k})$, the creation and annihilation
operators satisfy $ [ a_{\bf k}^{(\lambda)}  , a_{\bf
k'}^{\dagger(\lambda')}] = \delta_{\lambda\lambda'}\delta^3({\bf
k}-{\bf k'})$.

For  a fixed wave number $\bf k$ and a fixed polarization state
$\lambda$, the wave equation reduces to the second-order ordinary
differential equation  \cite{zhang,grish}
 \be
 h_k^{(\lambda)''}(\tau)
 +2\frac{a'}{a}h_k^{(\lambda)'} (\tau)+k^2h^{(\lambda)} _k =0,
 \ee
 where the prime denotes $d/d\tau$, and $k=|{\bf k}|$.  Since the equation of
 $h_{k}^{(\lambda)} (\tau) $
 for each polarization is the same, we denote $h_{k}^{(\lambda)}
(\tau) $ by $h_{k}(\tau) $ in below.

The power spectrum $h( k, \tau)$ of relic gravitational waves is
always defined as
 \be
 h(k,\tau) = \frac{4l_{Pl}}{\sqrt{\pi}}k|h_k(\tau)|.
 \ee
The spectral density parameter $\Omega_g(\nu)$ of gravitational
waves is defined through the relation $\rho_g/\rho_c=\int
\Omega_g(\nu)\frac{d\nu}{\nu}$, where $\rho_g$ is the energy
density of the gravitational waves and $\rho_c$ is the critical
energy density. One reads
 \be
 \Omega_g(\nu)=\frac{\pi^2}{3}h^2(k,\tau_H)(\frac{\nu}{\nu_H})^2,
 \ee
where $\tau_H$ is the conformal time of now and $\nu_H$ is the
Hubble frequency of now. Which can not exceed $10^{-6}$ as
required by the rate of the primordial nucleogenesis\cite{bbn}.

In the inflation models, RGW generated from the initial quantum
fluctuate during the inflationary stage. In general, in the
inflation stage, the scale factor has a form
 \be
 a(\tau)=l_0|\tau|^{1+\beta(\tau)}, ~~~-\infty<\tau<0,
 \ee

Normally, $\beta$ is a constant. If $\beta=-2$, the inflation is a
simple de Sitter expansion. But recently from observed results of
the primordial scalar power spectrum from WMAP\cite{map1}, one
finds the inflation is a very complicated process, here we assume
$\beta$ is a slowly evolving function with time $\tau$. For a
given wave number $k$, it crossed over the horizon at a time
$\tau_i$, i.e. when the wave length $\lambda_i = 2\pi a(\tau_i)/k$
is equal to $1/H(\tau_i)$, the Hubble radius at time $\tau_i$. Eq.
(7) yields $1/H(\tau_i) = l_0 |\tau_i|^{2+\beta}/|1+\beta|$, and
for the exact de Sitter expansion with $\beta =-2$ one has
$H(\tau_i)=l_0$. Note that a different $k$ corresponds to a
different time $\tau_i$. Now choose the initial condition of the
mode function $h_k(\tau)$ as\cite{zhang} \be |h_k(\tau_i)| =
\frac{1}{a(\tau_i)}. \ee Then the initial amplitude of the power
spectrum is
 \be
 h(k, \tau_i) = 8\sqrt{\pi}l_{pl}H(\tau_i),
 \ee
 From  $ \lambda_i = 1/H(\tau_i) $ it
follows that $ \frac{a'(\tau_i)}{a(\tau_i)} = \frac{k}{2\pi} $. So
the initial  amplitude of the power spectrum is
 \be
  h(k, \tau_i) =
Ak^{2+\beta(\tau_i)}=Ak^{2+\beta(-1/k)} ,
 \ee where the constant \be
A=8\sqrt{\pi}l_{pl}|1+\beta(\tau_i)|/l_0. \ee For the case of
$\beta=-2$, the initial spectrum is independent of $k$, this is
the so called scale independent initial power spectrum. Its
amplitude directly relate the inflation Hubble constant.

 The evolution of RGW in the following stage satisfies the two
 order differential equation (4). Here we only suppose that the
 Universe is always a decelerating expand. (from the research of SNIa, WMAP and LSS,
 one finds that now the Universe is accelerating expansion, we have researched its effect on the
 spectrum and found it was little\cite{zhang}. ) If the matter state equation satisfies
 $p=\omega \rho$, where $p$ and $\rho$ are the pressure and energy density
 of the matter respectively. ($\omega $ is a constant, for radiation, $\omega=1/3$; and for matter
 $\omega=0$), one can easily find that the scale factor has a form:
 \be
 a=b[\tau-\tau_1]^\alpha,
 \ee
where $\alpha=\frac{2}{3\omega+1}$, $b$ and $\tau_1$ are constant.
For the component of the Universe vary with time, we here think
the $\omega$ is varying slowly with time, this is a better way to
describe the evolution of the Universe than the method which
separated the Universe into several different evolution stages. So
the scale factor can be described in the following form:
 \be
 a=b(\tau)[\tau-\tau_1(\tau)]^{\alpha(\tau)},
 \ee
where $b(\tau)$, $\tau_1(\tau)$ and $\alpha(\tau)$ are all slowly
evolution functions with time. This way can be used to study all
kind of cosmic evolution models. Especially, it is useful for
studying the effect of the cosmological phase transition on the
RGW power spectrum.

In this model, when the wavelength of RGW is much larger than the
horizon, the amplitude of the spectrum keeps constant. Once the
gravitational waves enters the horizon, the spectrum will have a
damping along with the expansion the Universe. We assume that at a
certain short stage, the scale factor has a form of
 \be
 a=b'[\tau-\tau'_1]^{\alpha'},
 \ee
For $b'(\tau)$, $\tau'_1(\tau)$ and $\alpha'(\tau)$ vary slowly
with time, they can be assumed as constant in this stage. We study
the gravitational waves which enter the horizon in this stage. A
wave with wavenumber $k$, its amplitude satisfies\cite{zhang}
 \be
 h(\tau_H,k)=h(\tau_i,k)a_*(\tau)/a(\tau_H),
 \ee
where $a_*(\tau)$ is the scale factor when the wave $k$ just
enters horizon, and $a(\tau_H)$ is the now scale factor. From the
scale factor formula and the relation $k(\tau-\tau'_1)\simeq 1$
when wave enters horizon one find
 \be
 a_*(\tau)/a(\tau_H)=\frac{b(1/k)}{b_H}\frac{k_H^{\alpha(\tau_H)}}{k^\alpha},
 \ee
 where $b_H\equiv b(\tau_H)$. So we can get the spectrum
 \be
 h(k,\tau_H)=(A/b_H)b(1/k) k_H^{\alpha(\tau_H)} k^{\beta(-1/k)+2-\alpha(1/k)}
 =8\sqrt{\pi}l_{pl}H(\tau_i)(b(1/k)/b_H) k_H^{\alpha(\tau_H)}k^{-\alpha(1/k)},
 \ee
 For the slowly varying with time of $b(\tau)$, $\tau_1(\tau)$ and
 $\alpha(\tau)$, we can use this formula to estimate the RGW
 spectrum for all $k$. From this, one can also get the spectrum
 index
  \be
 n(k)\equiv d\ln h(k,\tau_H)/dk\simeq \beta(-1/k)+2-\alpha(1/k),
  \ee
So the matter state equation when the horizon of the Universe
equate  to $1/k$ is
 \be
 \omega(\tau)=\omega(1/k)=\frac{2}{3[\beta(-1/k)+2-n(k)]}-1,
 \ee
In general, $\beta+2$ is nearly zero for inflation, so this
formula directly relates the RGW spectrum index and matter state
equation in the Universe. For the waves with wavelength long
enough to have not reenter the horizon until now, they have the
power spectrum of their initial condition
  \be
  h(k, \tau_H)=h(k, \tau_i) = 8\sqrt{\pi}l_{pl}H(\tau_i)=Ak^{2+\beta(-1/k)},
  \ee
  and
  \be
 n(k)\simeq \beta(-1/k)+2.
  \ee

These are the approximate formulae, which can evaluate the RGW
power spectrum (If one want to know the exact spectrum, it must
depend on the numerical calculation or the complicated exact
calculation), but it is enough for knowing the physical
information which included in the RGW power spectrum. From formula
(20), one can find these spectrum keeps its initial value, which
reflects the physics of the inflation. Its amplitude is exactly
the Hubble parameter of the inflation, its change with $k$ is also
the change of Hubble parameter with time. So from which we can
know not only the energy scale of the inflation which closely
relates the inflation physics origin, but also the inflation
evolution information, which is also very important for the very
damping of the CMB anisotropies power spectrum in the very large
scale, and the large running of index of the primordial scalar
perturbation observed by WMAP\cite{map1,map2}, some people think
that it reflects the complicated evolution of early inflation.

For the waves with larger frequency, the spectrum formula (17)
also includes the evolution information of the Universe after
inflation. Its amplitude is dependent on the all evolution process
of Universe after cosmic horizon scale became larger than the
wavelength. Especially the spectrum index $n_s$, which directly
relates to the matter state equation when the cosmic horizon scale
equate to the wavelength, and the variety of the spectrum index
with wavelength is also the variety of the matter state equation
in the Universe with time. Actually, we can only detect the RGW
power spectrum in several narrow frequency span, for example the
laser interference explorers\cite{ligo}, which is sensitive for
the wave with frequency of $10^{-3}\sim 10^2$Hz, the method using
the millisecond pulsars\cite{mp}, which is sensitive at
$10^{-9}\sim10^{-7}$Hz, the method by researching the anisotropy
and polarization of CMB\cite{cmb}, which is sensitive at
$10^{-18}\sim10^{-16}$Hz. So even if we have detected RGW at some
frequency, we also can't know clearly the inflation energy scale
unless we know all information of Universe after the inflation
stage, but we can research the matter state equation in a special
stage of the Universe by the index of the power spectrum, which
directly relates the matter state equation $\omega$ at the time
$\tau=1/k$ by formula (19). So I think the simple RGW power
spectrum in all frequency can reflect the all evolution
information of the Universe since inflation, and it can not be
contaminated by other information as matter perturbation, the
interaction between the matters. So it is a very clear fossil for
study the cosmic evolution, and be a useful complementarity for
another fossil$-$CMB, especially for studying of the Universe
stage before the CMB photon decoupled.


\end{document}